\documentclass[preprint,12pt]{emulateapj}
\def\lsim{\lower.5ex\hbox{$\; \buildrel < \over \sim \;$}}
\def\gsim{\lower.5ex\hbox{$\; \buildrel > \over \sim \;$}}
\def\abeq{\lower.7ex\hbox{$\; \buildrel \sim \over - \;$}}

\def\t{\ifmmode {\tau} \else $\tau$ \fi}

\def\ref{\noindent \hangafter=1 \hangindent=0.7 truecm}

\def\cm{\ifmmode {\rm cm}^{-1} \else cm$^{-1}$ \fi}
\def\s{\ifmmode {\rm s}^{-1} \else s$^{-1}$ \fi}
\def\cc{\ifmmode {\rm cm}^{-3} \else cm$^{-3}$ \fi}
\def\cs{\ifmmode {\rm cm}^{-2} \else cm$^{-2}$ \fi}
\def\g{\ifmmode \gamma \else $\gammreview$\fi}
\def\G{\ifmmode \Gamma \else $\Gamma$\fi}

\def\kms{\ifmmode {\rm km\ s}^{-1} \else km s$^{-1}$\fi}

\usepackage{graphicx}

\begin{document}

\title{Deep Multiwaveband Observations of the 
Jets of  0208--512 and  1202--262}

\author{Eric S. Perlman\altaffilmark{1,2}, Markos
Georganopoulos\altaffilmark{2,3},
Herman L. Marshall\altaffilmark{4}, 
Daniel A. Schwartz\altaffilmark{5}, C. A. Padgett\altaffilmark{3},
Jonathan Gelbord\altaffilmark{6,7}, J. E. J.
Lovell\altaffilmark{8,9}, Diana M. Worrall\altaffilmark{10}, Mark
Birkinshaw\altaffilmark{10}, David W. Murphy\altaffilmark{11}, 
David L. Jauncey\altaffilmark{9}}

\altaffiltext{1}{Physics \& Space Sciences Department, Florida Institute of
Technology, 150 W. University Blvd., Melbourne, FL  32901, USA; eperlman@fit.edu}

\altaffiltext{2}{Department of Physics, Joint Center for Astrophysics,
University of Maryland-Baltimore County, 1000 Hilltop Circle, Baltimore, MD
21250, USA; georgano@umbc.edu, c13@umbc.edu}

\altaffiltext{3}{NASA's Goddard Space Flight Center, Mail Code 660, Greenbelt, 
MD, 20771, USA}

\altaffiltext{4}{Kavli Institute for Astrophysics and Space Research,
Massachusetts Institute of Technology, 77 Massachusetts Avenue, 
Cambridge, MA 02139, USA; hermanm@space.mit.edu}

\altaffiltext{5}{Smithsonian Astrophysical Observatory, 60 Garden Street,
Cambridge, MA 02138, USA;das@head-cfa.harvard.edu}

\altaffiltext{6}{Department of Physics, University of Durham, Science
Laboratories, South Road, Durham DH1 3LE, UK; j.m.gelbord@durham.ac.uk}

\altaffiltext{7}{Current Address: Department of Astronomy and Astrophysics, 525 Davey Laboratory, The
Pennsylvania State University, University Park, PA  16802}

\altaffiltext{8}{School of Mathematics \& Physics, Private Bag 21, University
of Tasmania, Hobart TAS 7001, Australia; Jim.Lovell@utas.edu.au}

\altaffiltext{9}{CSIRO Australia Telescope National Facility,  PO Box 76,
Epping NSW 1710, Australia; david.jauncey@csiro.au}

\altaffiltext{10}{H. H. Wills Physics Laboratory, University of Bristol, Tyndall Avenue,
Bristol BS8 1TL, UK; d.worrall@bristol.ac.uk, mark.birkinshaw@bristol.ac.uk}

\altaffiltext{11}{Jet Propulsion Laboratory, 4800 Oak Grove Drive, Pasadena, CA
91109, USA; dwm@sgra.jpl.nasa.gov}

\begin{abstract}

We present deep {\it HST, Chandra, VLA} and  {\it ATCA} images of the jets of
PKS  0208--512 and  PKS 1202--262, which were found in a {\it Chandra} survey of a
flux-limited sample of flat-spectrum radio quasars with jets (see
Marshall et al., 2005). We discuss in detail their X-ray morphologies and
spectra.  We find optical 
emission from one knot in the jet of PKS 1202--262 and two regions in the jet
of  PKS 0208--512.  The X-ray emission of both jets is most consistent with
external Comptonization of cosmic microwave background photons by particles
within the jet, while the optical emission is most consistent with the
synchrotron process.  We model the emission from the jet in this context and
discuss implications for jet emission models, including magnetic field and
beaming parameters. 

\end{abstract}

\maketitle
\eject

\section{Introduction}

Relativistic jets appear to be a common result of accretion onto compact
objects.  The jets of active galactic nuclei (AGN) include objects with a wide
range of luminosities and sizes.  In the most powerful sources -- quasars and
Fanaroff-Riley type II (FR II, Fanaroff \& Riley 1974)  radio galaxies -- the jets
terminate  at the hot spots,  compact bright regions hundreds of kpc away from
the nucleus of the host galaxy, where the jet flow collides  with the 
intergalactic medium  (IGM), inflating the radio lobes.  AGN jets are
enormously powerful, with a total bolometric power output that is often
comparable to or greater than that from the host galaxy, and a kinetic energy
flux that can be comparable to the AGN's bolometric luminosity (Rawlings \&
Saunders 1991). 

Until the last decade, almost all progress towards understanding the physics of
jets had come either from numerical modeling or multi-frequency radio mapping. 
However, the pace of discovery has accelerated greatly in the past decade with
{\it HST} and {\it Chandra} observations.  One of the first observations by {\it
Chandra} is a perfect illustration:  the target, PKS 0637--752, was a bright
radio-loud quasar. It was believed that this source would be unresolved in the
X-rays, and therefore ideally suited to focus the telescope. Instead, we saw a
beautiful X-ray jet well over 10 arcseconds long (Chartas et al. 2000, Schwartz
et al. 2000), with morphology similar to that seen in the radio.   Deeper
multi-band observations of this jet have since been done to constrain the nature
of its emission and also its matter/energy content  (Georganopoulos et al. 2005,
Uchiyama et al. 2005, Mehta et al. 2009).  Indeed, since the launch of the two
Great Observatories, the number of extended, arcsecond-scale jets known to emit
in the optical and X-rays has increased about ten-fold, from less than 5 to more
than 50, including members of every luminosity and morphological class
of radio-loud AGN.  

For jets, multiwaveband observations give physical constraints that cannot be 
gained in any other way.  In the bands where synchrotron radiation is the likely
emission mechanism (for the most part, radio through optical), one can combine
images to  derive information regarding particle
acceleration, jet orientation, kinematics and dynamics, electron spectrum and
other information.  The X-ray emission mechanism can be either synchrotron (with
very short particle lifetimes) or inverse-Compton in nature (for recent
reviews, see Harris \& Krawczynski 2006, Worrall 2009).  In either case, the X-ray emission
gives a new set of constraints on jet physics, because it represents
either (in the case of synchrotron radiation) the most energetic  particles 
($\gamma \sim 10^6-10^8$) in the jet, which must be accelerated {\it in situ},
or (in the case of inverse-Compton emission) the very lowest energy ($\gamma
\sim$ few - hundreds) particles, a population that controls the jet's
matter/energy budget yet cannot be probed by any other observations, 
and which still must be linked to the low-frequency radio
emission.  Combining information in multiple bands can then give unique
information on the magnetic field and beaming parameters as well as the
matter/energy budget of the jet.

Following the discovery of bright X-ray emission from the jet in  PKS 0637--752, 
we embarked on a survey (Marshall
et al. 2005, hereafter Paper I) to assess the rate of occurrence and properties of
X-ray jets among the population of quasars.  Our initial goals were to assess
the level of detectable X-ray fluxes from radio-bright jets, to locate good
targets for detailed imaging and spectral followup studies, and where possible
to test models of the X-ray emission by measuring the broad-band, spatially
resolved,  spectral energy distributions (SED) of jets from the radio through
the optical to the X-ray band.  The survey has been largely successful in
achieving these  aims, thanks to a relatively high success rate of about 60\%
of the first 39 observed (Marshall et al., 2011).  

We have used the survey data ($\sim 5$ ks {\it Chandra} observations) along with
other multiwavelength data  (including both longer {\it Chandra}  observations
as well as deep observations in other bands) to study the spectral energy
distributions and jet physics of several of the jets in our sample (Gelbord et
al. 2005; Schwartz et al. 2006a (hereafter Paper II), 2006b, 2007; Godfrey et
al. 2009).   Here we discuss relatively deep (35-55 ks) {\it Chandra} and {\it
HST} (2--3 orbits) observations of PKS B0208--512 and PKS B1202--262 (hereafter
0208--512 and 1202--262 respectively), two of the X-ray brightest jets found in
our survey.  Both had 40-115 counts detected in the initial {\it Chandra} survey
observation,  which detected multiple jet emission regions in the X-rays as well
as make an initial assessment of the jet physics (Paper II). The goal of these
deeper observations was to examine these conditions in greater detail, assess
the jet's morphology and spectral energy distribution in multiple bands, and
constrain the X-ray spectrum and emission mechanism.

0208--512 and 1202--262 were first discovered in the Parkes Survey of Radio Sources
(Bolton et al. 1964).  0208--512 was first identified by  Peterson et al. (1979) as a 
quasar at $z=1.003$, and is a powerful X-ray and Gamma-ray emitter.  It was
one of the  first extragalactic sources detected by both EGRET (Bertsch et al.
1993) and COMPTEL  (Blom et al. 1995), and has highly unusual gamma-ray
properties, being one of the  few blazars characterized by flares at MeV
energies, rather than in the GeV or TeV  domains (Stacy et al. 1996, Blom et
al. 1996).  1202--262 was 
identified as a quasar by Wills, Wills \& Douglas (1973) and Peterson et al.
(1979).  It is at a  redshift of $z=0.789$. 

In \S 2 we discuss the observations and data reduction procedures, while in \S
3, we discuss results from these observations.  In \S 4, we present a
discussion of physical constraints that can be gained from these observations. 
Finally, in \S 5, we sum up our conclusions. Henceforth, we use a flat, accelerating
cosmology, with $H_0 = 71 {\rm km ~s^{-1} ~ Mpc^{-1}} $ (for consistency with Paper II), 
$\Omega_M=0.27$ and  $\Omega_\Lambda=0.73$.

\section{Observations and Data Reduction}

\subsection {{\it Chandra} Observations}

Deep {\it Chandra} observations were obtained for both of our target sources, 
using the ACIS-S in FAINT mode.  {\it Chandra} observed 
0208--512 on 20 February 2004 for a total integration time of 53.66 ks, using a
standard 1/8 subarray and a frame time of 0.4 seconds.  The {\it Chandra}
observations of  
1202--262 were obtained 26 November 2004 for a total
integration time of 39.17 ks, using a standard 1/4 subarray and a frame time
of 0.8 seconds.  The
choices of subarray and frame time were driven by the core fluxes (Paper I), 
with the goal of
avoiding pileup greater than 10\%.  Roll angle ranges were requested to keep
the jet more than 30$^\circ$ from a readout streak.

\begin{figure}

\plotone{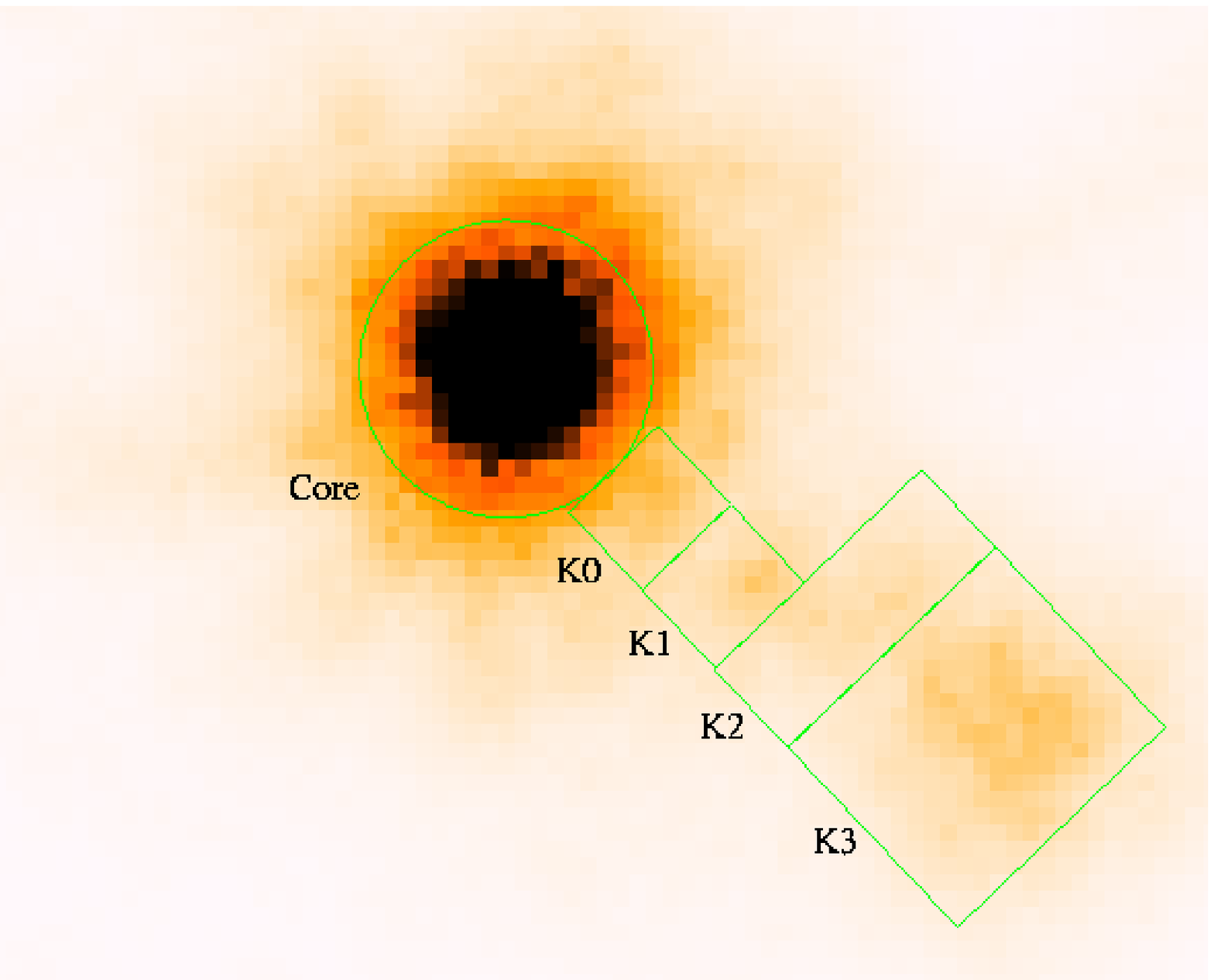}

\caption[]{Our deep {\it Chandra} observations of   0208--512.  Outlined in green are the core and
four jet regions used for photometry.    This image was sub-sampled to a resolution of 0.2 times the native 
ACIS-S pixel size (i.e., 0.0984$''$/pix) and adaptively smoothed with a minimum of 10 
counts per ellipse.  Three knot regions are clearly resolved
from the  nucleus (K1, K2, K3), while a fourth region (K0) that is seen in the
F814W HST  observations is possibly resolved from the nucleus.  
}

\plotone{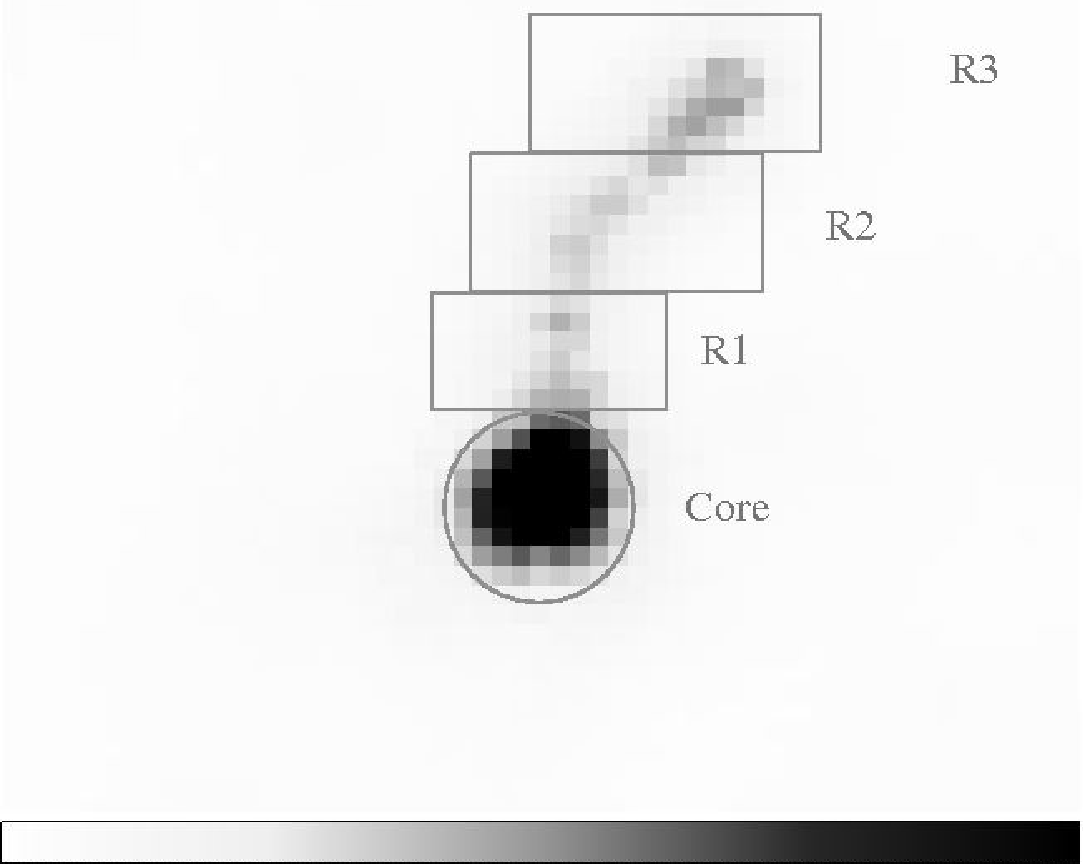}

\caption[]{Our deep {\it Chandra} observations of  1202--262.  
Outlined in green are the core and three jet regions used for
photometry.  This image was sub-sampled to a resolution of 0.5 times the 
native ACIS-S pixel size (i.e., 0.246$''$/pix) and adaptively  smoothed with a 
minimum of 10 counts per ellipse.  The morphology of this jet is quite smooth,
although a significant brightening is seen towards the terminus of the X-ray
bright region.  No change in surface brightness is seen as the jet bends gently
through about 50 degrees. 
} 

\end{figure}

The images were reduced in CIAO using standard recipes.  This included the
derandomization of photon positions within pixels, destreaking,  filtering out
flare-affected portions and bad pixels, as well as resampling images to 0.2 and
0.5 times the native ACIS pixel size (respectively 0\farcs092 and 0\farcs246 per pixel).  When
the data filtering was completed, the effective exposure times for the  two
observations were  48.67 ks for 0208--512 and 30.01ks for 1202--262. All
photons between 0.3-10 keV were used for both image and spectral analysis.   The
resulting {\it Chandra} image of 0208--512 is shown in Figure 1, while the
{\it Chandra} image of 1202--262 is shown in Figure 2.   


\subsection{{\it HST} Observations}

We also obtained {\it HST} observations for both sources.  We used the ACS/WFC
and obtained images in both the F814W and F475W bands.  We requested roll 
angles to keep the jet at least 30 degrees from a diffraction spike. Our
observing strategy was optimized both for long integrations as well as to obtain
a high-quality PSF for the central QSO.  This involved a combination of both long exposures (on
which the central pixels are saturated) and short exposures, with the latter serving
to fix the properties of the quasar PSF.  We CR-SPLIT
and  dithered each observation by integral pixel amounts to minimize the impact
of bad pixels and cosmic rays. In order to minimize the readout time we also
limited the WFC field of view to a 1024 $\times$ 1024 section of chip 1 (point
position WFC1-1K).    Our {\it HST} observations of  0208--512 took place on
26-27 May 2004 for three orbits.  The total exposure time in the F814W band was
3240s for the deep exposures and 340s for the shallow exposures, while in the
F475W band our integration times were 3114s for the deep exposures and 140s for
the shallow exposures.  The {\it HST} observations of  1202--262 took place
on 9 June 2004 for two orbits.  The total integration time in the F475W band was
1980s for the deep exposure and 100s for the shallow exposure, while in the
F814W band the total integration times were 2088s for the deep exposure and 100s
for the shallow exposure. 

One of our objects (0208--512) was also observed on 10 July 2004 for 2
orbits by the HST/ACS in the same bands by another team (led by F. Tavecchio), 
using a substantially simpler strategy.  In particular, they did not obtain
short exposures to measure the PSF, nor did they dither to eliminate bad
pixels.  Moreover, as  described in their paper (Tavecchio et al. 2007), in
those data the jet fell on a diffraction spike because the roll angle was not
specified. Because Tavecchio et al. also ended up using our data for their
analysis, we will refer to their paper when discussing any differences between
our findings and theirs for this source.

All HST data were reduced in IRAF and PyRAF using standard recipes.  The data
were drizzled onto a common grid using MultiDrizzle (Koekemoer et al. 2002 and
references therein).  As
part of the drizzling process, the images were rotated to a north-up, east-left
configuration. Since no sub-pixel drizzling was done, we did not 
subsample, hence leaving the data at the native ACS/WFC pixel size of 0\farcs05
per pixel.  All observations were used to obtain the final summed image, with
weights awarded according to their exposure time.  We also used MultiDrizzle to
sum the shorter exposures to obtain the QSO's fluxes and compare the PSF with
that generated by TinyTim (see next paragraph).

\begin{figure}
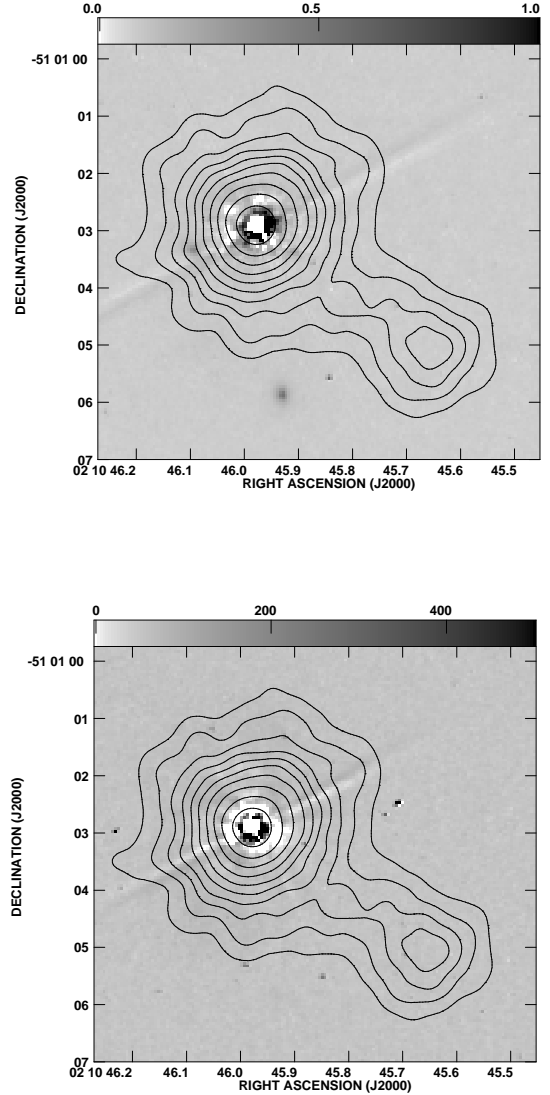


\plotone{pks0208_new_ch_f814w.eps}
\plotone{pks0208_new_ch_f475w.eps}

\caption[]{Our HST images of 0208--512 (greyscales) with {\it Chandra} X-ray
contours  overlaid.  At top, we show the F814W image, while at bottom we show
the F475W image. The Chandra image has been convolved with a $0.3''$ 
Gaussian.  A
scaled, TinyTim PSF has been used to subtract out the quasar point source in
the optical.  We have convolved the F814W image with a $0.2''$ 
Gaussian to better
show  the possible optical counterparts to knots K0 and K3, which are
particularly faint. Some evidence of oversubtraction is seen at the center of
the quasar; however, this is unavoidable given that the quasar is saturated and
given the general problem of charge bleed.  The contours are at (2, 4, 6, 8 ...) photons 
per pixel. See text for details.}

\end{figure} 

TinyTim simulations (Krist \& Burrows 1994, Suchkov \& Krist 1998, Krist \& Hook
2004) were obtained for PSF subtraction in both bands.  For the TinyTim
simulations we assumed an optical spectrum of the form $F_\nu \propto
\nu^{-1}$.  Previous experience with TinyTim has shown that the PSF  shape is
not heavily dependent on spectral slope.  Separate TinyTim simulations were
performed for each observation.  The most difficult part of the PSF subtraction
was correctly normalizing and rotating the PSF.  This necessitated independent
rotation of  each of the two axes to north-up (as the two axes of the WFC
detector are not completely orthogonal on the sky), as well as iteratively
looking at residuals and attempting to minimize the diffraction spikes. 
Importantly, because the charge 'bleed' in ACS data is linear, the total flux of
a saturated object is preserved even when heavy saturation is observed. Thus, in
these data, we optimized the PSF models for the subtraction of the outer
isophotes, comparing to the short exposures for the overall normalization.  
Because the long exposures were saturated, this also inevitably led to negatives
in the central pixels; however, given the small residuals in other places plus
the relatively smooth off-jet isophotes, we believe the result is reliable.

\begin{figure} 

\plotone{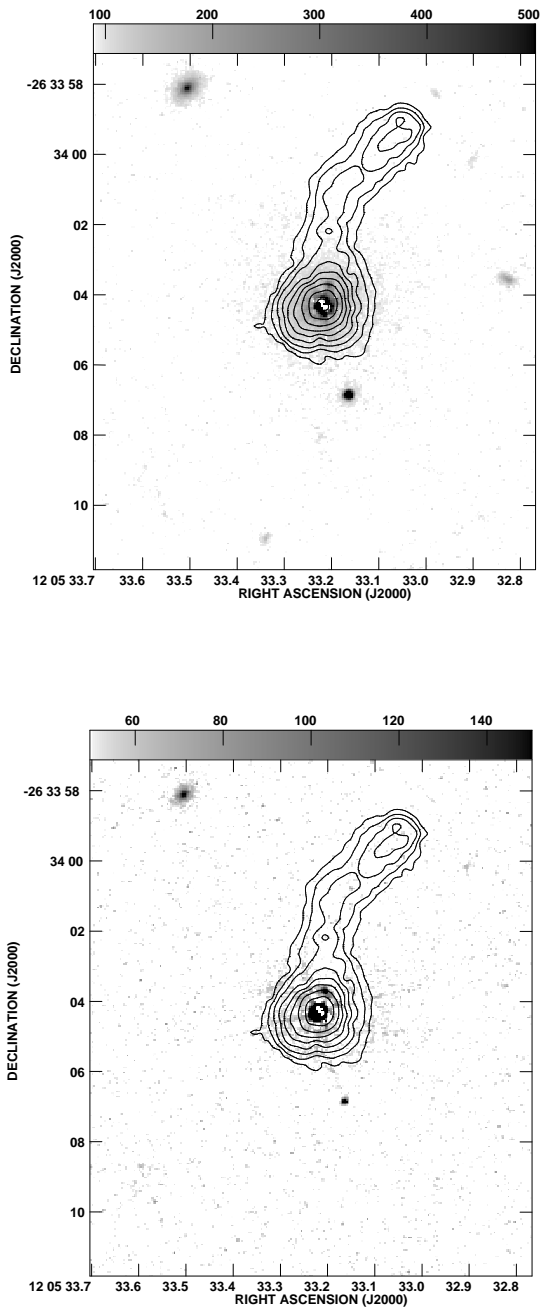}
\plotone{PKS1202_F475W_NEW.eps}

\caption[]{Our HST images of  1202--262 (greyscales), with Chandra X-ray
contours overlaid.  The pixellation and smoothing is as in the previous
figures. At top, we show the F814W image, and at bottom, we show the F475W
image.  The contours are at (2, 4, 6, 8 ...) photons 
per pixel. In both, a TinyTim PSF has been used to subtract out optical emission
from the quasar.  As with  0208--512's HST images, there is some evidence of
oversubtraction, but the problem is less severe for this object since its
quasar point source is less bright. Both images show an object $0.4''$ NNW
of the quasar.  This object is quite blue
(see text for discussion) and likely associated with a jet knot.}

\end{figure}


\subsection{Radio Observations}

The radio observations used in this paper were obtained at both the Australia
Telescope Compact Array (ATCA) and NRAO Very Large Array (VLA).  In Paper I we
presented ATCA observations of both sources at 8.6 GHz.   In addition to those
data we also obtained 4.8 GHz VLA observations of 1202$-$262 as well as 20.1 GHz
ATCA observations of both sources.  

The 4.8 GHz VLA data were obtained in the A configuration on 5 November 2000 as
part of project AM0672. The 8.6 GHz ATCA data were obtained in May and September
2000 for 1202$-$262 and February 2002 for 0208$-$512. Observations were made in
two array configurations, 6~km and 1.5~km to provide high angular resolution
imaging ($\sim$1.2 arcsec) as well as good sensitivity to extended structures.
The 20.1 GHz data from the ATCA presented here were collected as part of a
high-angular resolution radio-wavelength followup, complementary to the deeper
{\it Chandra} and {\it HST} observations.  These new ATCA observations were made
with the array  in a 6 km (6C)  configuration, providing an angular resolution
of ~0.5 arcsec FWHM, well matched to the {\it Chandra} PSF.  The 20.1 GHz
observations
of  0208$-$512 and 1202$-$262 were carried out on 2004 May 8 and 2004 May 13
respectively.  

Both sources were observed in full polarization at a bandwidth of 128 MHz.  The
sources were typically observed in groups of two or three with scans of 10
minutes duration interleaved over a 12 hour period.   At 8.6 GHz,  0208$-$512
was observed for a total of 310 minutes and  1202$-$262 was observed for 440
minutes. At 20.1 GHz, 0208$-$512 was observed for a total of 250 minutes and
1202$-$262 was observed for 290 minutes.  The  primary calibrator  1934$-$638
was observed and flux densities of 2.842~Jy and 0.923~Jy were assumed at 8.64
and 20.1 GHz respectively.   Short $\sim 1$ min scans of nearby compact
secondary calibrators were scheduled for the two program sources to provide
initial phase and gain calibration.

The ATCA data were calibrated in MIRIAD (Sault, Teuben \& Wright 1995) using the
standard reduction path for continuum data and, imaging was carried out with
Difmap (Shepherd 1997), where phase and amplitude self-calibration were
applied.  The images shown here were made using uniform weighting and restored
with a circular beam of equal area to the FWHM of the true synthesized beam. 
The VLA data were reduced using standard procedures in AIPS.

\begin{figure}

\plotone{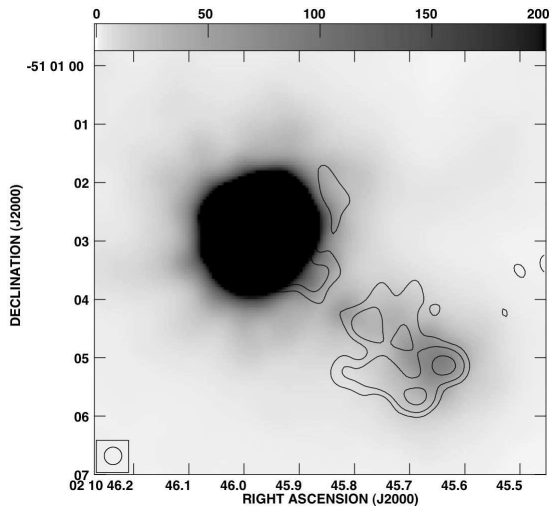}

\caption[]{Our {\it Chandra} image of  0208--512 (greyscale, with smoothing and 
pixellation as in Figure 3) with radio contours overlaid.  The radio contours are taken from 
our new 20.1 GHz ATCA data and have a restoring beam size of 0.44 arcsec. The contours are at 
(1, 2, 4, 8 ...) $\times $ 0.5 mJy/beam.  Note that
the radio and X-ray morphologies are somewhat different; in particular the
maxima of K1 and K2 are not at the same  locations, and the southeastern part
of knot K3 appears somewhat brighter in the radio (relative to the northwestern part) than in the X-rays.  See
text for discussion.} 

\end{figure}

\begin{figure}

\plotone{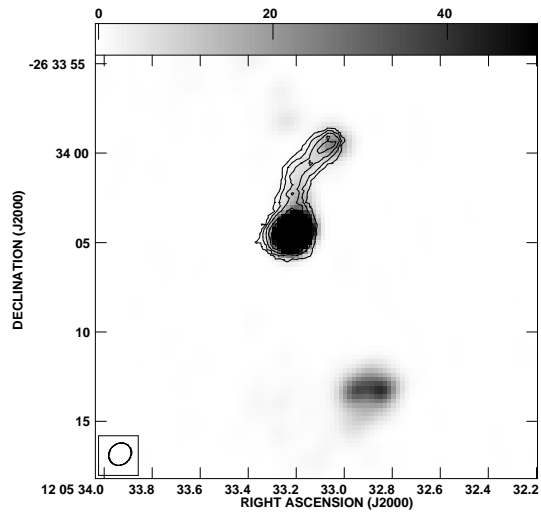}

\caption[]{Our ATCA 8.6 GHz image of  1202--262 (greyscale), with  {\it Chandra}
X-ray contours overlaid.  As can be seen, the X-ray emission from the jet  has
a morphology similar to that seen in the radio.  
No X-ray emission is seen from the region beyond the sharp
bend,  or from the counterjet side. }

\end{figure}

\subsection{Cross-Registration}

Images were registered to one another assuming identical nuclear positions in
all bands. As the absolute positional information in HST images is set by the
guide star  system and is typically only accurate to $1''$, the radio images
were used as the fiducial, adhering to the usual IAU standard. All other images
were therefore registered to the VLA image. Following this, the $1 \sigma$
errors in the positions from the HST images should be $\pm 0.2''$ in RA and DEC
(see e.g., Deutsch 1999), while the absolute errors in the X-ray image are $\pm
0.4''$\footnote{See the {\it Chandra} Science Thread on astrometry, 
http://cxc.harvard.edu/cal/ASPEC/celmon.}, relative to either the radio or
optical.  The {\it relative} errors, however, should be closer to $0.1''$, as
long as the assumption is correct that the quasar is the brightest source in
each band.  Armed with this information, we were then able to  produce overlays
of optical, radio and X-ray images.    

In Figure 3, we show the deep {\it HST} images of  0208--512, while in Figure 4,
we show the deep {\it HST} images of  1202--262.  Both figures show contours
made from the {\it Chandra}  images overlaid.  In Figure 5 and 6 respectively,
we show  overlays of radio contours onto the {\it Chandra} image of  0208--512
and 1202--262.   In Figures 7 and 8, we show radio/optical overlays of each
object.  Finally, in Figure 9, we  show the run of jet brightness with distance
for both objects, in both the 20 GHz radio  images and the X-ray images. To
produce the plots in Figure 9 we used the  {\it FUNTOOLS} ''counts in regions''
task, to produce radial profiles for each quasar plus jet and the quasar only.  
This allowed us to produce PSF-subtracted profiles, useful for  revealing
structure in the inner 1-2 arcseconds.

\begin{figure}

\plotone{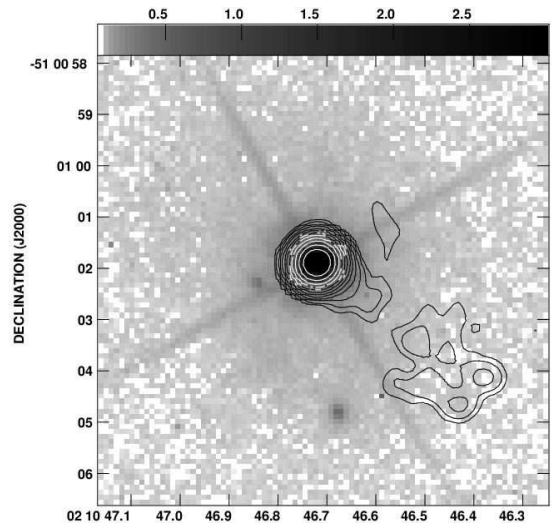}

\caption[]{Our F814W HST image of  0208--512 (greyscale) with  radio
contours from our 20 GHz ATCA image overlaid.  This figure shows the likely
link between the optical sources within the jet contours and jet components K0
and K3.  See text for discussion.}

\end{figure}

\begin{figure*}

\plotone{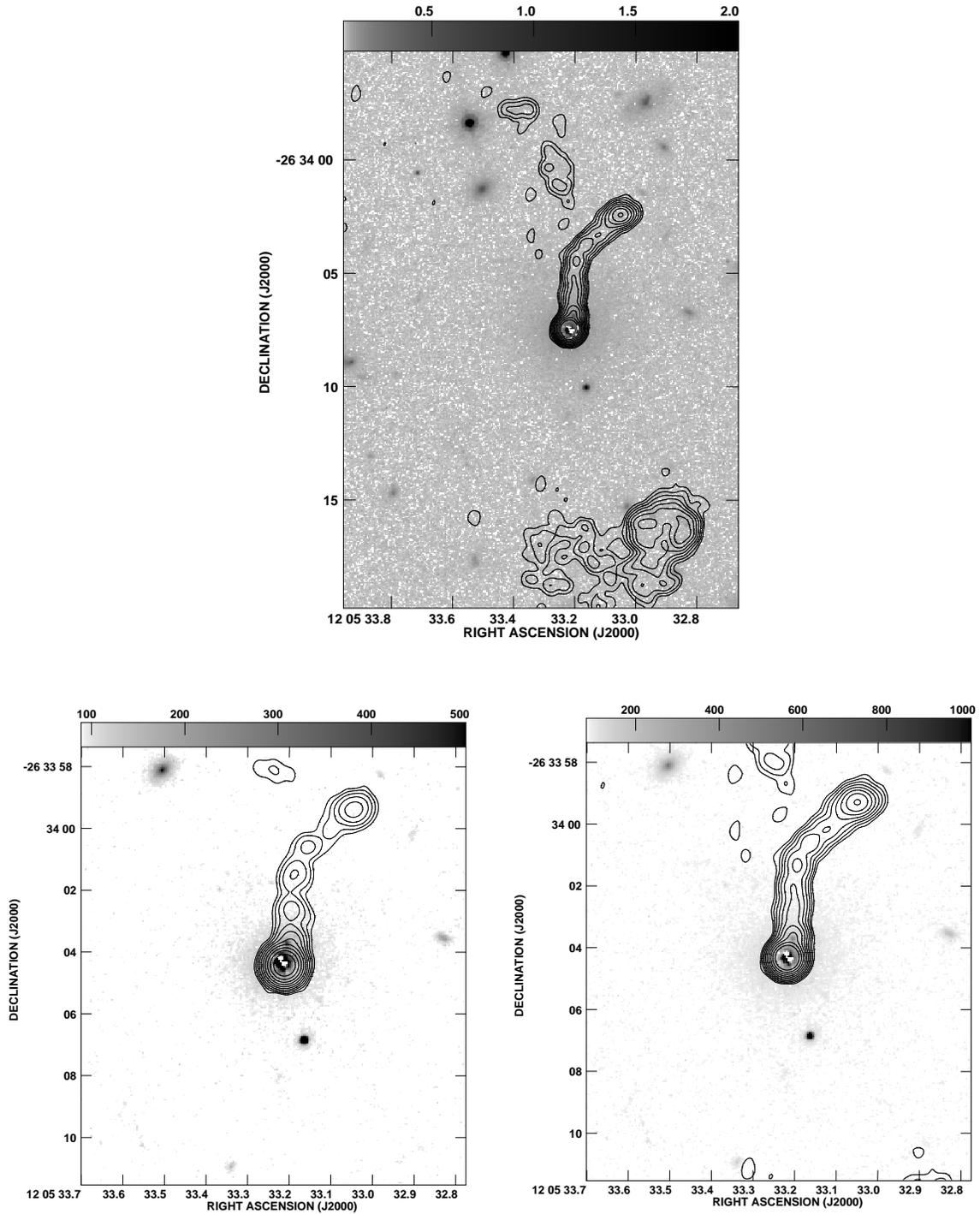}

\caption[]{Three views of  1202--262 showing the HST F814W image (greyscale)
overlaid with radio contours.  At top, we show a broader view, with contours
from the  4.8 GHz VLA A-array image.  As can be seen, no optical sources can be
firmly associated with any jet structure, with the exception of the source 
$0.4''$
north of the quasar.  At bottom, we zoom in on the X-ray bright region of the
jet, using the same region plotted in previous figures.  The left panel shows
the 20 GHz ATCA contours, while the right-hand panel shows the same 4.8 GHz VLA
A-array image shown in the top panel.  The 20 GHz ATCA image has a restoring beam 
size of 0.59 arcsec.  These plots cement the identification
of the optical source at $0.4''$ with a jet knot.  In addition, we see that the 
morphology of the jet changes quite significantly with frequency in the radio
band, with the higher-frequency map showing a much knottier jet that is quite
different from what is seen in X-rays.  The smoother morphology seen at lower
frequency is much more similar to the {\it Chandra} image.}

\end{figure*}

\begin{figure}

\plotone{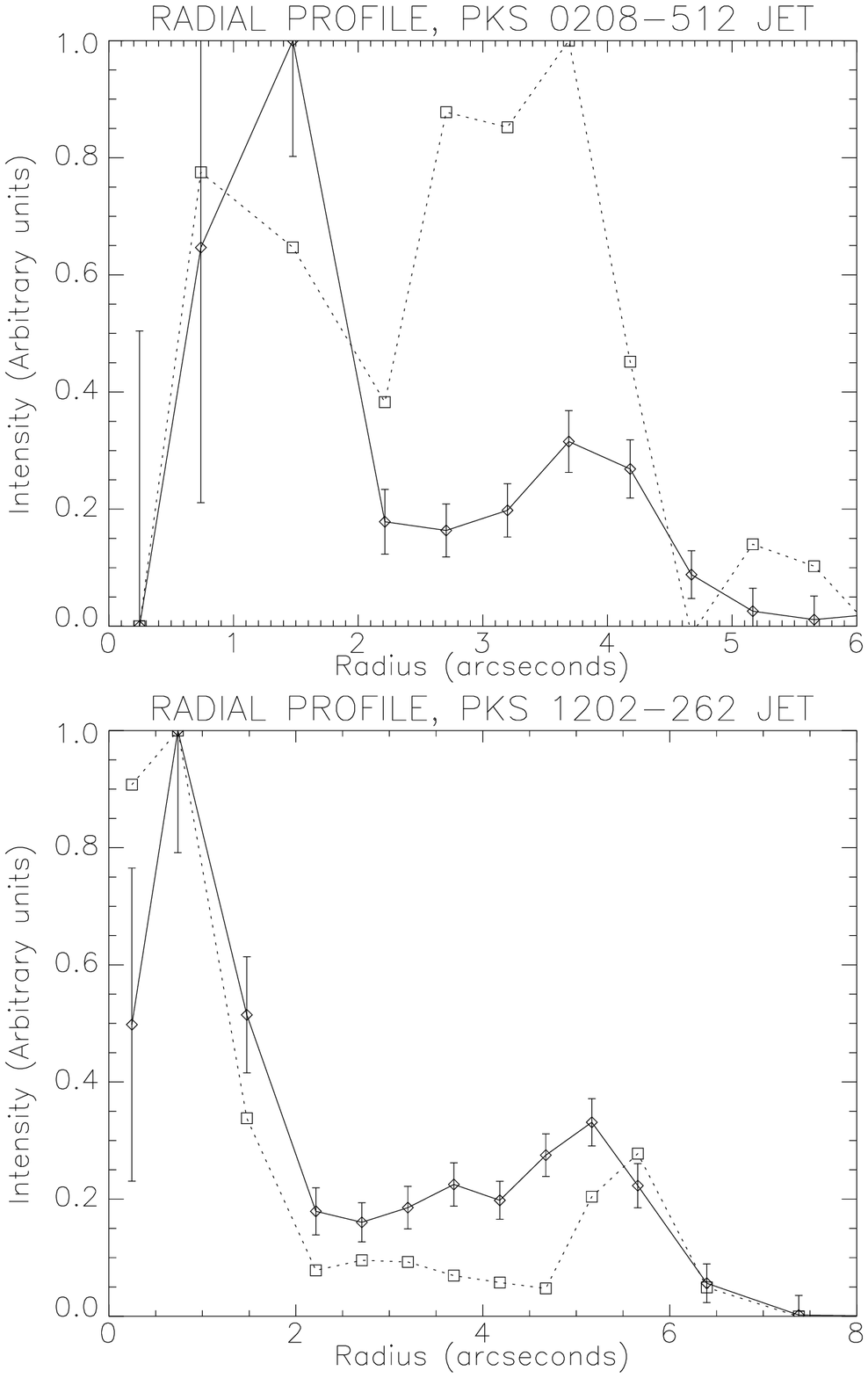}

\caption[]{Radial profiles for both the  0208--512 and  1202--262 jet.  The solid
line indicates the X-ray flux while the dotted line indicates the radio
profiles.  Note the significant differences between the radio and X-ray
morphologies of both jets.  See text for discussion.}

\end{figure}

\section{Results}

\subsection{PKS B0208--512}

The radio structure of 0208--512 is complex, 
and extends $5''$ SW of the core, with the innermost parts significantly more
compact than the rather diffuse outer component.   There is possible evidence of
a 90$^\circ$ bend in the 8.6 GHz ATCA maps given in Paper I; however, the
higher-resolution 20 GHz image (contours shown in Figure 5) shows the bend only 
at low significance.  VLBI  and VSOP observations (Tingay et al. 1996, Shen et
al. 1998, Tingay et al. 2002) show  a milliarcsecond-scale jet at a position angle 
similar to that seen on arcsecond scales.
Given the lack of lower-frequency radio data for this source at subarcsecond 
resolution, we did not attempt to make a radio spectral index map for  0208--512.
The X-ray structure of this source, as described by Paper II,
has two, possibly three emission regions, which were labeled R1, R2 and R3 (the
last labelled with a question mark due to the detection of only seven counts) 
in that paper.  As shown in Figures 1, 3, 5 and 7,  however, our deeper
observation shows the X-ray, radio and optical structure of this source to be
somewhat more complex.  We identify a total of four knot regions in the jet,
which we have labelled K0, K1, K2 and K3, in increasing order of distance from
the nucleus.  There is also a possible extension of the nuclear source seen in the
F475W image (Figure 3) only; however, it is not fully resolved, and could also
be an artifact of the undersampling of the PSF combined with the location of the
quasar within the central pixel.  We therefore do not attempt to measure
quantities associated with this source.

The relationship of  K0-K3 to the emission regions described by Paper II is complex. 
Region K0 lies closer to (distance 0\farcs8 from) the nucleus than any of the
regions pointed out by Paper II, and is not well resolved from the nucleus in
the X-rays.  It is, however, easily resolved from the nucleus in radio and
appears to have a faint optical counterpart (Figures 3 and 5). When the point
source is subtracted from both the radio and X-ray images in a radial profile
(Figure 9), K0 is seen more clearly at distances of 0.7-1.5 arcseconds from 
the nucleus.  The data contain tempting hints of a significant morphology 
difference between the radio and X-ray in K0, as the flux peak is located closer to the nucleus in the radio than the X-ray.  However, the errors in the X-ray point source subtraction are large enough to make this not definitive.  K1
corresponds roughly to the region labelled R1 by Paper II;  hence the name,
chosen to minimize confusion.  This region is the faintest in radio and X-rays.
There is faint optical emission in this region; however, there is
nothing that stands out above the level of the galaxy.  Therefore, we do not
believe this jet region has significant optical emission connected  with it.  

The region labelled R2 in Paper II is seen to split into two fairly distinct
maxima, which we have here called K2 and K3. K2 is seen easily in the
radio as well but any optical emission lies below our detectability threshold. 
It should be noted, however, that the sole radio maximum in the region of K1 and
K2 actually lies in between these knots.  It is therefore unclear whether these
represent knots in the physical sense.  K3 appears to
have a rather complex structure.  In the radio, two distinct peaks are seen
along with a component on the south side that appears to extend back towards
the nucleus.  The X-ray emission appears to come from the entire region
but is brighter on the north side.  The comparison of radio and X-ray flux 
profiles (Fig. 9)
reveals very different morphologies in these two bands, with the radio maximum
in between K1 and K2 not corresponding to a significant flux increase in the X-rays,
while the radio flux decreases quite a bit faster in the downstream part of K3 than
that of the X-rays.  There does appear to be a 27th magnitude
optical source at the position of  the southern radio flux maximum of K3, which
we believe is associated with the jet; however no significant optical flux is
detected at the position of K3's northern maximum.   All of the optical
counterparts to jet emission regions are seen only in the F814W image; in the
F475W data they are below the detection limit.  We do not detect X-ray or
optical emission from the region called R3 by Paper II, despite the much deeper
{\it Chandra} exposure obtained here.

The finding of optical emission within the jet in at least two regions,  differs
significantly from the  results of Tavecchio et al. (2007), who report no
optical emission within the  jet of 0208--512.  Those authors, however, did
not do a point-source subtraction of the {\it HST} data.  Given the brightness
of the optical AGN it would be quite easy to miss faint optical sources if 
PSF subtraction were not done, particularly in the inner 2 arcseconds, where
region K0 and the possible extension seen in F814W are found.  Tavecchio et al.
also did not do a detailed comparison to radio data which reveal the good
positional coincidence between the optical emission in K0 and the first radio
knot.  

From the {\it Chandra} data we extracted spectra separately for the core and the 
jet, using a background region defined by a circular annulus that
excluded both the core and jet.  For the core region (which also included K0), 
we obtained  F(0.5-2 keV) = 1.85
$\times 10^{-12} ~{\rm erg ~cm^{-2} ~s^{-1}}$ and  F(2-10 keV) = 3.59 $\times
10^{-12} ~{\rm erg ~cm^{-2} ~s^{-1}}$.  Its spectrum was well fit by a
power-law index of $\Gamma = 1.72 \pm 0.05$ and a  N(H) = $3.0_{-0.8}^{+0.9}
\times 10^{20}$ ~cm$^{-2}$, the latter being comparable to the Galactic value of
$2.9 \times 10^{20}$ ~cm$^{-2}$.  The reduced
$\chi^2$  was 0.956 and the probability of a null hypothesis was 0.67.  These
parameters are somewhat different from  those reported in Paper I for the 5 ks
observations in Cycle 3; in particular, they indicate variability at about the
20\% level (in the sense that the core was brighter at the time of the Cycle 3
observations).  However, the 90\% confidence error intervals of the spectral indices 
for the two
spectral fits just overlap, so the spectrum of the jet did not change
significantly.  The spectral index we observe is consistent with $\Gamma=1.7$,
as reported for the cumulative X-ray  emission of this source in Sambruna
(1997) and Tavecchio et al. (2002) from ASCA  and SAX data respectively.  

For the jet, we used a region that included all of regions K1-K3.  
We obtained  F(0.5-2 keV) = 5.67 $\times 10^{-14} {\rm ~erg
~cm^{-2} ~s^{-1}}$ and F(2-10 keV) = 1.01 $\times 10^{-13} {\rm ~erg ~cm^{-2}
~s^{-1}}$.  Its spectrum was well fit by a power-law index of
$\Gamma=1.69_{-0.35}^{+0.36}$, with N(H) being fixed at the Galactic value.  
This represents the spectrum of the entire jet, for which we observe about 400
counts during the observations.  


We extracted radio, optical and X-ray flux information from the regions shown
on Figure 1 for each knot.   These are shown in Table 1. 
That Table also shows $2\sigma$ upper limits for the regions not seen in
either F814W or F475W. 

\begin{deluxetable}{lcccc}[t]
\tablewidth{-0pt}
\tablecaption{Fluxes of Components in PKS 0208--512}
\tablehead{
\colhead {Region} & \colhead{F(20 GHz)} & 
\colhead{F(F814W)} & \colhead{F(F475W)} & \colhead{F(1 keV)} \\
& \colhead {(mJy)} & \multicolumn{3}{c}{(nJy)} }

\startdata

Core & 2540 & $2.46 \times 10^6$ & $1.22 \times 10^6$ & 554 \\
K0   & 2.70 & 77 & $<9.9^{\rm a}$ & 2.301\\
K1   & 1.35 & 69 & $<9.9^{\rm a}$ & 0.801 \\
K2   & 2.85 & $<14$ & $<9.9^{\rm a}$ & 0.842 \\
K3   & 11.93 & 31 & $<9.9^{\rm a}$ & 4.088 \\
\enddata

\tablenotetext{a}{All upper limits are 2 $\sigma$.}

\end{deluxetable}

\subsection{PKS B1202--262}
	      
\begin{deluxetable}{lcccc}[t]
\tablewidth{-0pt}
\tablecaption{Fluxes of Components in PKS 1202--262}
\tablehead{
\colhead {Region} & \colhead{F(20 GHz)} & 
\colhead{F(F814W)} & \colhead{F(F475W)} & \colhead{F(1 keV)} \\
& \colhead {(mJy)} & \multicolumn{3}{c}{(nJy)} }

\startdata
Core & 553.1 & 3.58 $\times 10^6$ & $3.83 \times 10^6$ & 136 \\
R0   & 11.13 & $4220$ & $1770$ &... $^{\rm b}$ \\  
R1   & 10.05 & $<26^{\rm a}$ & $<18^{\rm a}$ & 7.854 \\
R2   & 8.25  & $<26^{\rm a}$ & $<18^{\rm a}$ & 9.431 \\
R3   & 15.24 & $<26^{\rm a}$ & $<18^{\rm a}$ & 11.288 \\
CJ   & 47.08 & $<26^{\rm a}$ & $<18^{\rm a}$ & $<0.30$ 
	    
\enddata

\tablenotetext{a}{All upper limits are 2 $\sigma$.}
\tablenotetext{b}{Not resolved from the core.  See discussion in \S 3.2.}

\end{deluxetable}

We see X-ray emission from all along the northern jet (Figures 4), with the
exception of the faint, bent radio structure beyond the 120$^\circ$ bend (Figure
8).  The jet shows a very smooth X-ray morphology, without significant ``knots",
making its structure more similar to that seen in C-band than at higher radio
frequencies.  This can be seen clearly in the comparison of the X-ray and radio 
jet profiles (Figure 9). This is more consistent with inverse-Compton than
synchrotron emission, as most of the high-power jets where synchrotron is the
suspected  X-ray emission mechanism (3C 273,  PKS 1136--135) show a much
'knottier' X-ray  morphology.  Significant  brightening of the X-ray jet is seen
at its terminus (what we are calling the  radio hotspot near the sharp bend). 
Interestingly, the radio flux peak in this region clearly peaks further from the
nucleus than does the X-ray  flux (Figure 9).  We have labelled the X-ray 
resolved regions of the jet R1-R3, as shown in Figure 2. 

The radio structure of this source is characterized by a bent jet that extends
for  about $8''$ from the core, ending in a hotspot, as well as a
southern  (counterjet-side) lobe situated about $10''$ away from the nucleus.  The jet
side also has a  fainter, bent extension to the northeast 
in the radio that comes off the jet at
roughly a 120 degree angle $2''$ from  its end but is not seen in the X-rays. 
The relationship of this fainter
structure to the main jet is not known; however,  it is unlikely to be
serendipitous.   This source has one of the brightest X-ray jets in our sample
(and was the brightest  selected from our Cycle 3 {\it Chandra} observations). 
It is remarkable in that the X-ray  flux in the jet is about 10\% of that in
the core.   Our radio data (Figures 8, 10) shows all regions of the jet.  The
jet is considerably knottier in the high-frequency images than it is at lower
frequencies, with the region beyond the northeastern extension not seen at 
frequencies greater than 4.8 GHz.  This is consistent with steep-spectrum, 
extended jet regions.  The hotspot in the counterjet is also considerably fainter at 
high frequencies.

For this jet, the existence of both VLA 4.8 GHz and ATCA 20.1 GHz data, with
very  nearly identical resolution, makes it possible to make a radio spectral
index map. We show this map in Figure 10, overlaid with contours from the 20.1
GHz VLA image. Both the nucleus as well as the innermost ($0.4''$) knot show a
flat spectrum ($\alpha_r \sim 0$), whereas the rest of the main jet has
$\alpha_r$ between 0.5 and 1.0.  Both of these are reasonably typical.  There is
some correlation between flux and spectral index within the jet, particularly in
the two brightest radio knots.

A 20th magnitude optical
knot is seen at 0\farcs4 from the nucleus, along the same PA as the jet.  We
have labelled this region R0.   No
other optical emission is seen associated with the jet,  to a limit of about
28.5 mag.   While the one optical source observed (R0) is too close to the core to be
resolved clearly in the {\it Chandra} image, there is a distortion of the 
flux contours in that direction, and moreover, when the X-ray point source is 
subtracted from the radial profile (Figure 10), a clear excess is seen corresponding
tot the location of a radio knot (Figure 8) seen on higher frequency, higher
resolution radio maps.  Thus  there can be  little
doubt about the association of this optical emission with the jet. The optical
knot is seen in both the F814W and F475W image and is quite blue, with an
F475W-F814W color of 0.7 mag.  By comparison, the typical elliptical galaxy
would be far redder, with an F475W-F814W color of  $\sim 2-3$ mag.


We extracted radio, optical and X-ray flux information from the regions shown
on Figure 2 for each knot.   These are shown in Table 2. 
That Table also shows $2\sigma$ upper limits for the regions not seen in
either F814W or F475W.

\begin{figure}

\plotone{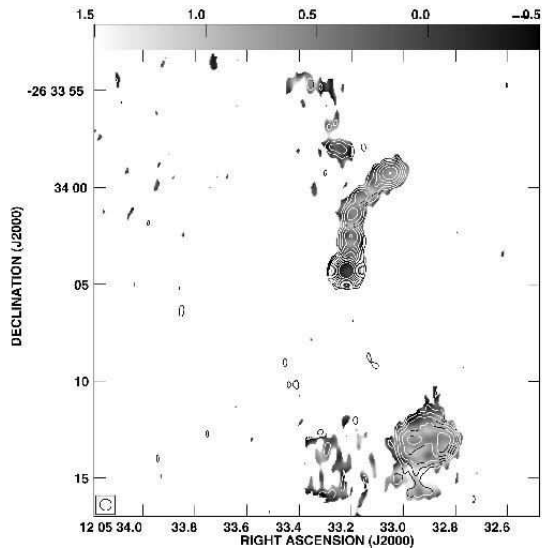}

\caption[]{Radio spectral index map of  1202--262 between 4.8 and 20.1 GHz, 
overlaid with flux contours taken from the 4.8 GHz VLA image.  See text for 
discussion.}

\end{figure}

We extracted X-ray spectra separately for the core and jet, as well as for a background 
annular region centered on the source that excluded all source flux.  For the
core region (which also includes R0, which is unresolved from the core in X-rays), 
we obtained 
F(0.5-2 keV) = 4.55 $\times 10^{-12} {\rm ~erg ~cm^{-2} ~s^{-1}}$ and 
F(2-10 keV) = 1.44 $\times 10^{-12} {\rm ~erg ~cm^{-2} ~s^{-1}}$.  The core
was well fit by a single power-law spectrum with $\Gamma = 1.45 \pm 0.03$ and
N(H) fixed at the Galactic value of 7.08 $\times 10^{20} {\rm ~cm^{-2}}$.  The
reduced $\chi^2 = 0.954$ and the probability of a null hypothesis was 0.68.  
These parameters indicate significant variability in the core as compared to
the Cycle 3 parameters given in Paper I; specifically, the 
core was about 30\% brighter in these Cycle 5 observations, and the spectral
index was also significantly flatter in these observations.  Unlike Paper I, we 
do not see any evidence for a narrow Fe K$\alpha$ fluorescence
line at a rest energy of 6.4 keV.  However, given the ample evidence for both
flux and spectral variability of the core between 2002-2004 we do not think
this refutes Paper I's assertion of the possible detection of 
that line in the Cycle 3 data.  

For the jet (which includes regions R1, R2 and R3), we obtained 
F(0.5-2 keV) = 5.90 $\times 10^{-14} {\rm ~erg ~cm^{-2} ~s^{-1}}$ and 
F(2-10 keV) = 1.74 $\times 10^{-13} {\rm ~erg ~cm^{-2} ~s^{-1}}$.  The 
jet was well fit by a power-law spectrum with $\Gamma = 1.50 \pm 0.10$ and a
N(H) fixed at the Galactic value.  The reduced $\chi^2$ for this fit was 
0.828, with a probability of the null hypothesis of 0.81.  


\section{Discussion}

We have presented new, deep images of the jets of  0208--512 and  
1202--262 in three bands:  the radio, optical and X-rays, using multiple
telescopes.  These data allow us to construct detailed models of these jets and
their emission processes, and comment on physical parameters and how (if) they
change with distance from the quasar core.  In addition,  as these sources are
among the brightest X-ray jet sources known, the results of such an analysis
has implications for our knowledge of the class as a whole, since the number of
sources where such an analysis has been done is still small. We have plotted in
Figure 11 the luminosity of various jet regions in
each source.  Here we
proceed to discuss the nature of each of the jet regions in terms of the most
popular jet emission models, under which the X-ray emission  seen by {\it
Chandra} is either synchrotron emission or CMB emission that has been
Comptonized by the synchrotron-emitting jet particles.  We discuss each of
these in turn.

\subsection{Synchrotron X-ray Emission}

One possibility is that the X-ray emission is due to  synchrotron emission,
which is the most likely mechanism in less powerful radio galaxy jets (e.g.,
Worrall et al. 2001, Perlman \& Wilson 2005, Harris \& Krawczynski 2006).  This model was 
applied to quasar jets by Dermer \& Atoyan (2002; see also Atoyan \& Dermer 
2004), in the context of a single
electron population under which the upturn in the  broadband spectrum would be
explained by the fact that the Klein-Nishina limit becomes relevant at these
energies, thus imposing a different energy  dependence to their
spectral aging.  That model cannot succeed for most of the bright X-ray quasar
jets, because it is unable to produce jets where the optical emission is well below 
the $\nu F_\nu$ level seen in the X-rays.  
As can be seen in Figure 11,
both of  our jets are in this regime, with the exception 
of the optically bright knot $0.4''$ north of the core of 1202--262, which is 
not clearly resolved.  However, the brightness of this knot's optical emission
makes its SED very different from the other knots (Figure 11). 

A more elaborate version of the synchrotron model holds that a second,
high-energy population of electrons is present (Schwartz et al. 2000, 
Hardcastle 2006, Harris
\& Krawczynski 2006, Jester et al. 2007).   The X-ray spectral indices we
observe in the jet ($\alpha = \Gamma - 1 = 0.7$ in the case of the jet of 
0208--512 and $\alpha = 0.5$ in the case of the jet of 1202--262) require
electron energy distributions (EEDs) that  locally would be a power law
$n(\gamma) \propto \gamma^{-p}$, with $p = 2 \alpha_x +1 = 2 - 2.4$.  However,
if these electrons are in the regime where their radiative aging is
not determined by the  Klein-Nishina cross-section, and if the cooling time is
significantly faster than their escape time, then the aceleration mechanism that
produced them is required to provide an EED of $n(\gamma) \propto \gamma ^{-1}$
to $\gamma^{-1.4}$.  
Note that adiabatic losses do not steepen the electron energy distribution
and thus result in similar spectral indices (Kirk et al. 2000)

Two-zone synchrotron models effectively double the number of
free parameters needed to model the broadband jet emission. Moreover, they
force one to consider different and physically distinct acceleration mechanisms
and zones for the electron populations producing the emissions seen in X-rays
and at lower frequencies.  In particular, the nature  of the optical ``valley"
seen in these jets requires that the X-ray emitting particles must be
accelerated at energies of at least $\sim 10$ TeV and then accelerated up to 
at least 100 TeV before they escape, necessarily to an environment of much 
lower magnetic field, so that they do not produce substantial optical-UV
synchrotron emission as they cool (see Georganopoulos et al. 2006 for details).
Therefore, we do not favor these models, although we cannot rule them out.

\subsection{Comptonized CMB X-ray Emission (EC/CMB)}

A more commonly invoked model is that the X-ray emission observed from these
jets is  due to inverse-Comptonization of the cosmic microwave background (the
so-called EC/CMB model, Celotti et al. 2001, Tavecchio et al. 2000). This model
has the  virtue of requiring only a single population of electrons, and in
addition takes advantage of a mandatory process to produce the high-energy
emission.  Central to this model is the requirement that the jet remain
relativistic out to  distances of many kiloparsecs from the quasar nucleus with
bulk Lorentz factor  $\Gamma\sim 10$. This is significantly larger than the
constraints form radio  jet to counter jet flux ratios (Arshakian \& Longair
2004) that require  $\beta=u/c \gtrsim 0.6$  ($\Gamma\gtrsim 1.25$).

The EC/CMB model requires a continuation of the EED to low Lorentz factors --
energies which currently cannot be observed in any other way.  Electrons at
these very low energies by necessity will dominate the overall energy budget of
the  jet (just because of their sheer numbers) and thus the models are also
constrained by the need to not violate the Eddington luminosity for the
supermassive black hole (see  Dermer \& Atoyan 2004 for the general argument and
Mehta et al. 2009 for  its detailed application on PKS  0637--752). To make use
of the large $\Gamma$,  jets have also to be well aligned to the line of sight
$\theta\sim 1/\Gamma$  and this poses additional constraints in the model 
because the actual size of  the jet becomes at least $\Gamma$ times the
projected size (Dermer \& Atoyan 2004).

  In the EC/CMB model the X-ray emission is related to very low-energy electrons
and thus we would naively expect to see a very smooth X-ray emission, outwardly
similar to what is observed at low radio frequencies.  This is definitely what
is observed in the jet of  1202--262, where the X-ray morphology is very smooth,
without significant knots (Figure 9), much more similar to  what is seen in the
4.8 GHz VLA map than to what we see at 20 GHz. However it  is hard to comment on
whether this is the case in  0208--512 given the smaller angular size of its
jet.  Knotty X-ray emission could also be expected in the EC/CMB model under two
scenarios.  The first of these is that the injection in the jet varies and 
that the X-ray knots correspond to
periods of  increased plasma injection that are now traveling downstream. 
Alternately, the well known ``sausage'' instability due to a mismatch in
pressure between the jet and external medium could create alternating
compressions and rarefactions.

In our previous work (in particular Paper II), we adopted an approach patterned 
after the work of Felten \& Morrison (1966), who showed that under the
assumption  that the X-ray emission from all the knots arises from
inverse-Compton scattering by the same power-law population of electrons which 
emit the synchrotron  radiation, the ratio of synchrotron to Compton power will
be equivalent to the ratio of the energy density of the magnetic field to that
of the target photons:

\begin{equation}
{S_{synch} \over S_{IC} } =  {{(2 \times 10^4 T)^{(3-p)/2} B_{\mu G}^{(1+p)/2}}
			   \over {8 \pi \rho}}, 
\end{equation}
	
{\noindent where $\rho = \Gamma^2 \rho_0 (1+z)^4$ is the apparent energy density of the 
CMB at a redshift $z$ in a frame moving at Lorentz factor $\Gamma$, $\rho_0 =
4.19\times 10^{-13} {\rm ~erg ~cm^{-3}}$ is the local CMB energy density, and the apparent
temperature of the CMB is $\delta T$, where $\delta = [\Gamma (1-\beta \cos
\theta)]^{-1}$ is the usual beaming parameter, and $T=2.728 (1+z)$ K is the 
CMB temperature corresponding to epoch $z$.   }

This approach makes two assumptions. First, the CMB  energy density in the comoving 
frame [$(4/3)U_{CMB, local}\Gamma^2$]  is treated as isotropic (this is stated in 
the discussion before eq. 3 of Schwartz et al. 2006).  And second, the beaming of 
synchrotron and EC/CMB is the same and for that reason no $\delta$ appears in the 
ratio.  However, neither of these is required by the physics and in fact both are problematic.

\subsection{Independently Solving for $B, \delta$}

We can make further progress on understanding the physical conditions in our
emitters by making use  of the formalism adopted by Georganopoulos, Kirk \&
Mastichiadis (2001)  for blazar  GeV emission.  We consider a blob of plasma
permeated by a magnetic field $B$,  moving relativistically with a bulk Lorentz
factor $\Gamma$ and velocity $u=\beta c$ at an angle $\theta$ to the observer's
line of sight.  In the frame of the blob, then, the electrons are  characterized
by an isotropic power-law density distribution, 

\begin{equation}
n^\prime(\gamma^\prime) = {{k}\over{4\pi}} \gamma^{\prime -p} P(\gamma_1, \gamma_2, \gamma^\prime),
\end{equation}

{\noindent where $\gamma^\prime$ is the Lorentz factor of the electron (in the 
blob's frame of reference), $k$ is a constant, and $P(\gamma_1, \gamma_2, 
\gamma^\prime)=1$ for $\gamma_1 \leq \gamma \leq \gamma_2$ and zero otherwise.  
Under the assumption that $\gamma^\prime >>\Gamma$, one can treat the electrons 
as a photon gas, so that we have a radio spectral index $\alpha=(p-1)/2$.  If we
make use of the Lorentz invariant 
quantity $n/\gamma^2$, the Lorentz factor $\gamma$ of an electron in the lab frame 
is then $\gamma=\delta \gamma^\prime$, where the Doppler factor 
$\delta=[\gamma(1-\beta\cos\theta)]^{-1}$.  Then, the electron density $n(\gamma)$ in the lab frame is }

\begin{equation}
n(\gamma,\mu) = n^\prime(\gamma^\prime)\Big({{\gamma}\over{\gamma^\prime}}\Big)^2 = 
{{k}\over{4\pi}} \delta^{2+p}\gamma^{-p} P(\gamma_1\delta,\gamma_2\delta,\gamma),
\end{equation}

{\noindent where $\mu=\cos\theta$.
Given that the effective volume $V_{\rm eff}$ of the blob in the lab frame is $V_{\rm eff}=V~\delta$  -- 
as shown in Georganopoulos et al. (2001) -- with $V$ being the blob's volume in its own frame, the 
energy distribution of the effective number of electrons $N_{\rm eff}(\gamma,\mu)$ is }

\begin{equation}
N_{\rm eff}(\gamma,\mu) = n(\gamma,\mu) V_{\rm eff} = 
{{kV}\over{4\pi}} \delta^{3+p}\gamma^{-p} P(\gamma_1\delta,\gamma_2 \delta,\gamma).  
\end{equation}

Using this,  we obtain for the external Compton luminosity:

\begin{equation}
L_{IC}(\nu)=c_2k\delta^{4+2\alpha} \nu^{-\alpha} 
\end{equation}

{\noindent for a moving blob.  Similarly, for a standing feature through which the plasma flows relativistically we would obtain}

\begin{equation}
L_{IC}(\nu)=c_2k\delta^{3+2\alpha} \nu^{-\alpha},
\end{equation}

{\noindent where the constant $c_2$ can be found in Mehta et al. (2009). 
Note that this equation  does not depend on the bulk Lorentz factor, but it does assume $\gamma_2>>\gamma_1$.
The next equation comes from the observed synchrotron luminosity:}

\begin{equation}
L_s(\nu)=c_1 k \delta^{3+\alpha} B^{1+\alpha}\nu^{-\alpha}
\end{equation}

{\noindent for a moving blob or}

\begin{equation}
L_s(\nu)=c_1 k \delta^{2+\alpha} B^{1+\alpha}\nu^{-\alpha}
\end{equation}

{\noindent for a standing  feature through which plasma flows relativistically,
and  constant $c_1$ can be found in Mehta et al. (2009).  We now have two equations 
and three unknowns, $B$, $k$, and $\delta$. Note that the bulk motion Lorentz factor 
does not enter our equations. }

To close this system of equations we require that the source is in the equipartition configuration. Following Worrall \& Birkinshaw
(2006) the magnetic field $B$ in equipartition conditions is related to the electron normalization $k$ through 
\begin{equation}
{B^2 \over 8\pi}={\alpha+1 \over 2} (\chi+1){k m_e c^2 (\gamma_1^{1-2\alpha}-\gamma_2^{1-2\alpha})\over V}, 
\end{equation}

{\noindent where $\chi$ is the ratio of cold to radiating particle energy density, $V$ is the source volume, and $\alpha$ is the radio spectral index.
 If  we make the assumption that $\alpha>1/2$, as justified by knot observations, and $\chi=1$, we obtain}

\begin{equation}
{B^2 \over 8\pi}=(\alpha+1){k m_e c^2 \gamma_1^{1-2\alpha}\over V}. 
\end{equation}

Now we have three equations (EC/CMB luminosity, Synchrotron luminosity, and
equipartition), and three unknowns, $B$, $k$, $\delta$. These may be solved to
uniquely determine them. The values of $B$ and $k$ fix the energy density in the
blob. A choice of $\theta$ provides a bulk Lorentz factor $\Gamma$ and vice
versa.  With a choice of  $\Gamma$ in hand, we can then calculate the jet power.

Note that the above equations give a somewhat different approach than our
previous work.  The differences  lie in the beaming parameters, which come from
the different expressions we have for the EC/CMB emission. Both approaches agree
on the equations for the synchrotron luminosity (eqs. (7) and (8)).

If one adopts the Georganopoulos et al. (2001) formalism that we use here,  then
using the results of equations (7) and (9) we obtain the following for the ratio
of the two luminosities $L_s$ and $L_{IC}(\nu)$:

\begin{equation}
{L_{s}(\nu)\over L_{IC}(\nu)}\propto \delta^{-(1+\alpha)} .
\end{equation}

Note that this ratio is independent of bulk $\Gamma$ or $\mu$.  Similar results
were derived by Dermer (1995), with an extra multiplication factor of $[(1 +
\beta)/(1+\mu)]^{(1+\alpha)}$, which, as discussed in Georganopoulos et al.
(2001), comes from the approximation that the seed photons in the frame of the
blob for inverse Compton scattering are coming from a direction opposite to the
direction of the blob velocity.  The Dermer (1995) equations are presented in a
form independent of the system of units by Worrall (2009).  The extra
multiplication factor is so close to unity for bulk speeds and angles to the
line of sight appropriate for quasar jets (and in particular for the sources
presented here), as to render the formalisms essentially identical.  Dermer \&
Atoyan (2004) and Jorstad \&  Marscher (2004) also used very similar formalisms
in their analysis. The fundamental advantage to this approach is that it allows
us to calculate $B$ and $\delta$ directly, with the Lorentz factor of the jet
$\Gamma$ then  resulting because of the choice of angle $\theta$.  This allows
us to find directly a  maximum viewing angle for the jet, $\theta_{max}$ (i.e.,
where $\delta = 2 \Gamma$).

\subsection{Application to our Data}

\begin{figure*}[t]
 
\plottwo{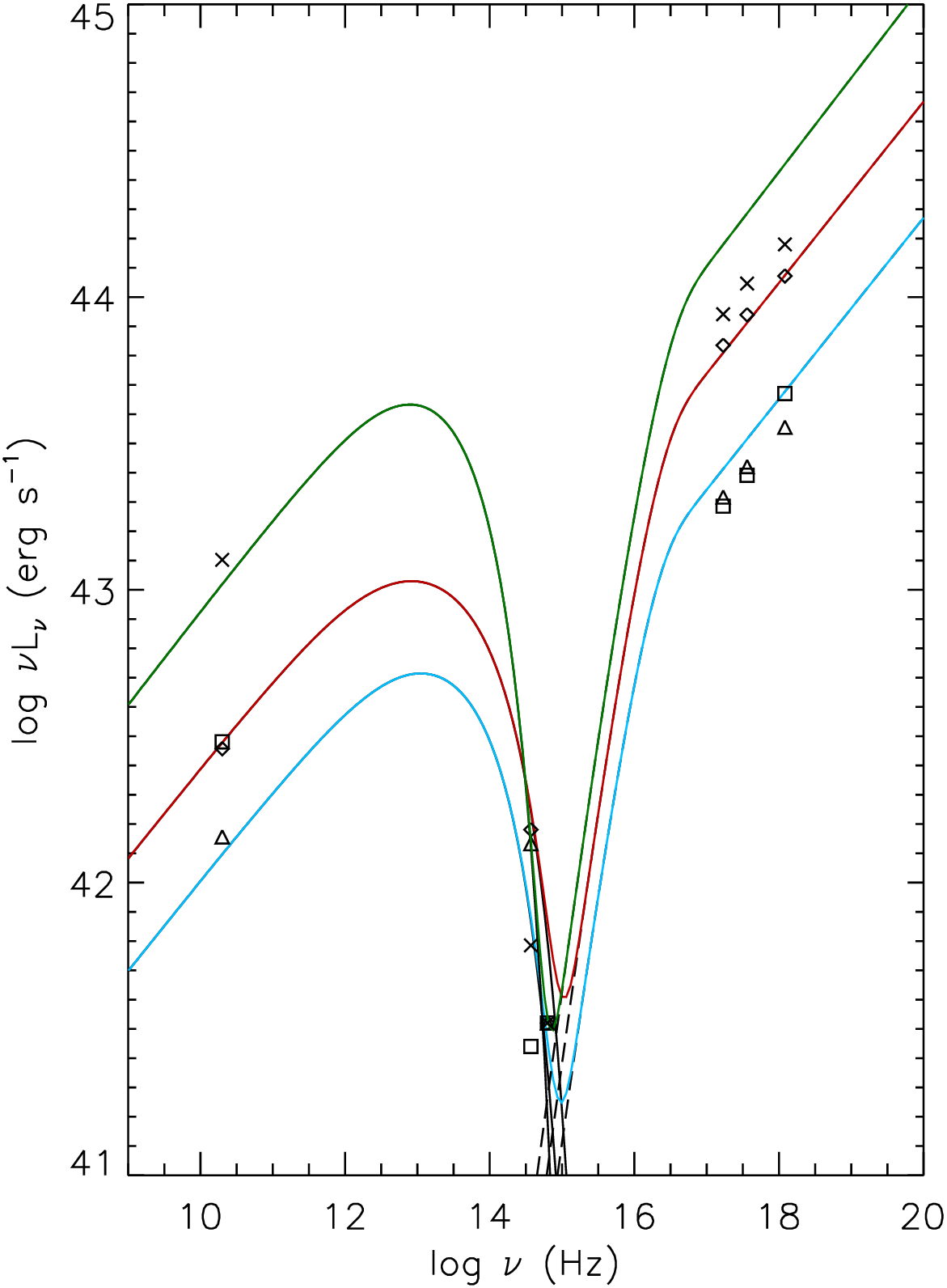}{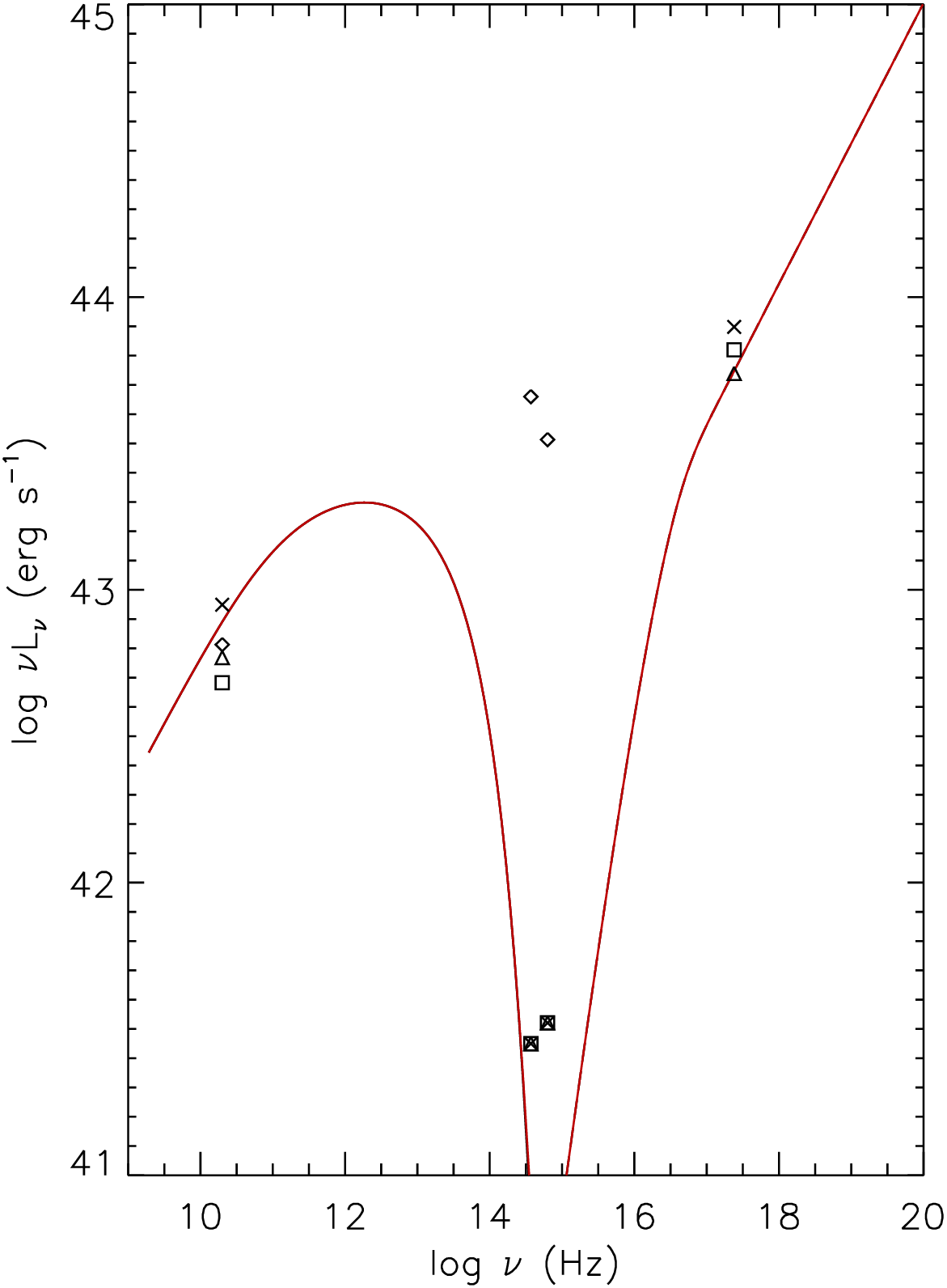}

\caption[]{Spectral energy distributions (SEDs) for the core plus four regions 
of the jet of  0208--512 (left) and  1202--262 (right).  For the jet of 0208--512, data points
for K0 are shown as diamonds, those for K1 are triangles, those for K2 are squares, and 
the data for K3 are X's.  For the jet of 1202--262, data points for R0 are diamonds, data 
for R1 are triangles, those for R2 are squares, and the data for R3 are X's.  With the exception
of R0 in the 1202--262 jet, all the F475W points are upper limits, as is the case for the F814W points for 0208--512 K2, 1202--262 R1, 1202--262 R2 and 1202--262 R3.  The most 
likely interpretation for the optical emission of the detected knots of both jets is the 
extreme tail of synchrotron emission, while the X-ray emission is most likely due to 
inverse-Comptonization of CMB photons (IC-CMB), as discussed in \S 4.2. We have 
overplotted Synchrotron/IC-CMB models of jet components, as detailed in \S 4.2.2.  For 
0208--512, we show models for regions K0 (red), K1 (blue) and K3 (green); a model is
not shown for K2 since it is essentially identical to that for K2.  For 1202, we show only 
the model for R1, as the models for R2 and R3 are essentially identical.  No model is 
fitted for R0 in the 1202 -262 jet as the knot is not fully resolved in X-rays, making it difficult to constrain the IC-CMB component.}

\end{figure*}

We now apply the method laid out in \S 4.3 to calculate physical parameters for
the jets of 0208--512 and  1202--262, under the assumption that the X-ray emission
from them arises via the EC/CMB scenario.
%
%
%
We adopt a single spectral index $\alpha = (p-1)/2$, equal to that observed in
the radio  (i.e., 0.69 for 0208--512 and 0.52 for 1202--262).    These spectral
indices are consistent  with those observed in the X-rays, as would be expected
under the EC/CMB model.
We adopt a uniform magnetic field
and assume the equipartition condition.  We
also assume that the emitting volume is a cylinder, with a filling factor of
unity and assume that the ratio of proton to electron energy  density is also 1.
All emitting  volumes are taken to be cylinders with the measured angular lengths,
and radii of 0.5 arcseconds.  We assume a moving blob for each feature with the 
given values of $\delta$, $B$, etc. (as opposed to a standing shock).  
The minimum particle energy was chosen as $\gamma_{min}=15$.
These assumptions result in the values laid out in Table 3. The resulting 
spectral energy distributions are plotted in Figure 11.  The typical error 
ranges on these parameters are $\sim 10\%$, so that, for example, the jet of 
PKS 1202--262 is consistent with a constant $\delta$ and $\theta$ for all 
components.  Note too that in estimating these parameters we assumed that the 
same volume is radiating for both the radio and X-ray, whereas in Figure 9 
and accompanying text we have noted differences between the X-ray and radio 
morphologies for both jets.
To calculate the jet kinetic power, one needs to select a viewing angle.  For
PKS 0208--512, we have chosen a value of $\theta = 3^\circ$ , which was the mean
angle to the line of  sight for the MOJAVE sample (Hovatta et al. 2009), whereas
for PKS 1202--262, which is required to be more highly beamed  ($\delta>20$ and
$\theta < 2.5^\circ$) we chose $\theta=2^\circ$.  The jet powers reported in 
Table 3, would increase by a factor of $m_p(p-2)/[m_e(p-1)\gamma_{min}]$ if
we assume instead that for every radiating electron there is a cold proton. 
The result would be an increase in jet power by 
a factor of $\approx 34$ for 0208--512 and $\approx 2$
for 1202--262. Even with this increase -- which corresponds to the limiting
case, under rough equipartition conditions, of the absence of positrons -- 
the jet powers remain smaller or at most 
comparable to the Eddington luminosity of a $10^9 M_\odot$ black hole.   
%
%
\begin{deluxetable}{ccccccc}    
\tablewidth{0pt}
\tablecaption{Properties of the X-Ray Jet Components }
\tablecolumns{7}
\tablehead{
\colhead{Object}& \colhead{Region} & \colhead{$B ~(\mu $G)} & 
\colhead{$\delta $} &\colhead{$\theta_{\rm max}$ (deg)}  & \colhead {$L_{com}^{\rm a}$}  &
\colhead{$P_{jet}^{\rm b}$}}

\startdata

0208--512 & K0 & 8.7 & 10.2    & 5.6 & 1.03  & 2.10 
 \\
		&	K1 & 8.8 & 8.3  & 6.9 & 1.06  & 1.36  
		\\
		&     K2 & 12.0 & 7.5  & 7.7 & 1.96 & 2.00  
		\\
		&     K3 & 13.9 & 9.5   & 6.0 & 2.65 & 4.60 
\\
1202--262 & R1 & 4.9   & 22.5 &2.5 & 0.31 & 3.84 
\\
		&    R2 & 4.2   & 24.8 & 2.3 & 0.23 & 3.99 
		\\
		&   R3  & 5.3   & 23.4 & 2.4 & 0.36 & 1.42 

\enddata

\tablenotetext{a}{Comoving luminosity, in units of $10^{44} ~ {\rm erg ~s^{-1}}$.}
\tablenotetext{b} {Kinetic power, in units of $10^{45} {\rm ~erg ~s^{-1}}$.}
 \end{deluxetable}
 
\section{Discussion}

We have analyzed deep {\it Chandra} and {\it HST} observations of the jets of 
PKS 0208--512 and PKS 1202--262.  Both jets can be seen in the X-ray images 
extending for several arcseconds from the quasar, and both exhibit at least one 
optically detected component.  The observed X-ray characteristics of both jets are 
consistent with the popular IC-CMB model but are difficult to explain under models 
where the X-ray emission is due to the synchrotron process.  In this context, we have 
presented a method to analyze the X-ray emission of jets and solve independently
for the beaming factor $\delta$ and magnetic field $B$.  


The two jets appear to illustrate two rather different ranges of parameter space.   It is
apparent that the jet of PKS 0208--512 requires higher magnetic fields and 
lower beaming parameters $\delta$, and
a greater angle to the line of sight than the jet of PKS 1202--262.
We find only weak evolution, if any, of the physical parameters with distance from the 
nucleus in both jets (Table 3).  In 0208--512 there is evidence for a consistent increase in 
magnetic field with increasing distance from the nucleus; however, a similar pattern is not
seen in 1202--262.
While there is some change in the Doppler factor the jet of PKS 0208--512, it is not systematic in
either one.
Changes in the Doppler factor would produce differences between the observed radio and X-ray morphology
(Georganopoulos \& Kazanas 2004), in the sense that in a gradually decelerating
jet we would expect X-ray knots to lead those seen in the radio, due to 
additional emission from the upstream Compton mechanism.  However, looking at 
Figure 9, it is interesting to note that there are significant radio/X-ray morphology differences in 
both jets.  As can be seen, the X-ray fluxes of several jet components do appear to be located closer
to the nucleus than the corresponding radio maxima.  This is in line with the predictions of the upstream
Compton scenario of Georganopoulos \& Kazanas (2004) but does not require changes in the jet speed
or trajectory with increasing distance along the jet.

\begin{acknowledgments}

ESP, CAP and MG acknowledge support from {\it Chandra} GO 
grants G02-3151D and G04-
5107X), HST (grant STGO-10002.01) and NASA (LTSA grants NAG5-9997 and
NNG05-GD63DG at UMBC and NNX07-AM17G at FIT). HLM was 
supported under NASA contract SAO SV1-61010 for the {\it Chandra} X-Ray Center (CXC). 
JMG was supported under {\it Chandra} grant GO2-3151A to MIT from the CXC. DAS was 
partially supported by {\it Chandra} grant G02-3151C to SAO from the CXC, and by
NASA contracts NAS8-39073 and NAS8-3060 to the CXC.

\end{acknowledgments}

\vfill\eject

 \end{document}